\documentclass[12pt]{article}

\input epsf
\usepackage{epsfig,graphicx,epsf}
\usepackage{amsmath,amsfonts,amssymb,amsbsy,textcomp}

\def\i{\mbox{i}}
\def\\i{\mbox{\scriptsize{i}}}

\title{On the shape of nucleons at high energies }
\author{M. L. Nekrasov\\
{\small\it 
Institute for High Energy Physics, NRC ``Kurchatov
Institute'',}  \vspace*{-4\baselineskip}\\
{\small\it Protvino 142281, Russia} }
\date{}
\begin{document}
\maketitle
\begin{abstract}
A scenario of evolution of the shape of nucleons with increasing energy
is described in the framework of extended parton model, which
consistently takes into account the transverse motion of the partons. At
the energy $E$ up to LHC, the nucleons have the form of a spheroid which
expands as $\sqrt{\ln \!E}$ in the transverse directions and grows
linearly in $E$ in the longitudinal direction. With a further increase
in the energy a mode of correlated behavior of the partons is
established, which stops the longitudinal growth. Simultaneously, the
expansion in the transverse directions changes to $\ln \!E$, and a
hollow mostly free of partons is formed inside the nucleons along the
central axis in the direction of their motion. Numerical estimates of
the corresponding parameters are obtained.
\end{abstract}

\section{Introduction}\label{sec1}

The spatial shape of hadrons is one of the intriguing problems in the
hadron physics. A closely related problem is the spatial region of
interaction between the hadrons in elastic collisions at high energies.
An independent solution to these problems would explain the behavior of
many characteristics of hadron scattering. Accordingly, the study of
these problems has been at the forefront of theoretical research for
many years
\cite{Gribov73009,Feinberg,Gribov65,Low,Schrempp,Gribov73,
Petrov-Ryutin,Nekrasov2015,Nekrasov2017,Blok,Petrov-Okorokov,Nekrasov}.

\begin{figure}[t]
\hbox{ \hspace*{55pt} 
       \epsfysize=0.45\textwidth \epsfbox{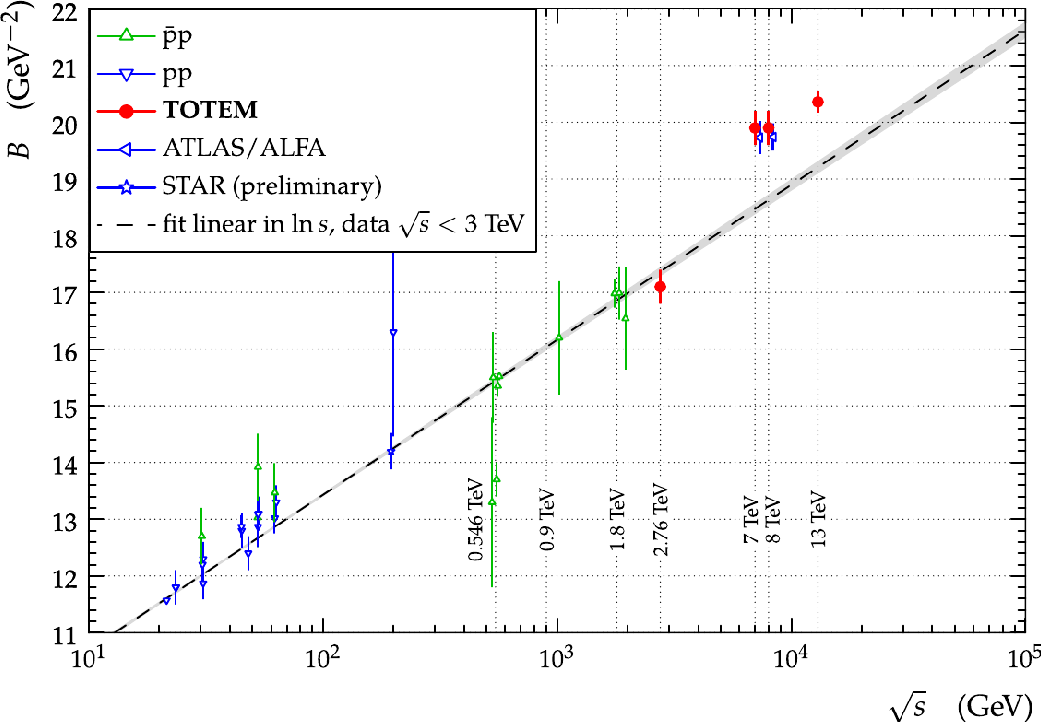}}
\caption{\small The energy dependence of the slope B of the diffraction
cone \cite{Antchev} } \label{Fig1}
\end{figure}

The main attention is usually paid to the transverse sizes of hadrons
and the region of their interaction. In the first approximation, they
can be judged by the total cross section. However, more accurate
information is provided by the slope parameter $B$ of the diffraction
cone in the elastic collisions. Namely, due to relation $B = \langle b^2
\rangle / 2$, which links $B$ with the mean square of the impact
parameter $b$, one can judge the transverse sizes of the interaction
region. The transverse sizes of the hadrons must be related to $\langle
b^2 \rangle$, as well. Experimental data on the nucleon scattering
indicate a growth of $B(s)$ as $\ln s$ up to the Tevatron energy
\cite{Amaldi}, and then an accelerated growth at the LHC energies
\cite{Antchev}, possibly as $\ln^2 \!s$ \cite{Schegelsky-Ryskin},
cf.~Fig.\ref{Fig1}. The total and elastic cross sections exhibit similar
behavior~\cite{Antchev}. 

In theory, such a behavior is predicted by Regge model. Really, at
moderately high energies it describes the elastic scattering of hadrons
via the one-pomeron exchange. This leads to $B = 2 \alpha'_0 \ln
(s/s_0)$, where $\alpha'_0$ is the slope of the pomeron trajectory. At
ultra-high energies the multi-pomeron exchanges lead to $B \propto \ln^2
\!s$ and the same behavior for the total cross section. The transition
to the multi-pomeron exchange mode is expected in the 2--7 TeV region
\cite{Schegelsky-Ryskin}. 

This behavior must be explained at the ``microscopic'' level of quarks
and gluons, as well. Unfortunately, the means of perturbative QCD
can hardly be useful in this case since the relevant processes occur at
extremely small transfers. However, as they occur at high energies, the
parton model is suitable for appropriate consideration. In that way the
result, which became classical, was obtained about the ``diffusion''
growth of the transverse sizes of hadrons as $\sqrt {\,\ln s}$
\cite{Gribov73}, which means the behavior $B(s) \propto \ln s$, observed
at the moderately high energies. In the analysis \cite{Gribov73},
hadrons were considered free-moving, which can be associated with their
elastic scattering in the limit $t=0$ (excluding electromagnetic
interactions). Recently an extension of this approach was proposed
\cite{Nekrasov}, which took into account in more detail the transverse
motion of the partons. The latter analysis predicted a faster growth of
the transverse sizes of hadrons in the transition to the ultra-high
energies, from $\sqrt{\,\ln s}$ to $\ln s$. This means the transition of
the behavior of $B(s)$ from $\ln s$ to $\ln^2 \!s$. 

Unfortunately, the situation with the longitudinal sizes of hadrons
remains uncertain. The point is that no experimental criteria for
estimating the longitudinal hadron sizes have been proposed so far.
Therefore, the reasoning on this topic is rather speculative. Actually,
two points of view currently dominate. The first one is associated with
the concept of a hadron at rest as a rigid ball or bag of approximately
spherical shape. Based on this, the fast hadron due to Lorentz
contraction should look like a highly flattened pancake with a
longitudinal size of the order of $1/E$, where $E$ is the energy of the
hadron. According to the other point of view \cite{Gribov73} the
longitudinal sizes are determined by the minimum momenta of the partons,
at which they still belong to the hadron (i.e.~separated from the
vacuum). From the latter point of view, the longitudinal size of the
hadron does not depend on the energy and is of the order of $1/\mu$,
where $\mu$ is the soft scale in the nucleon.

However, there is a third point of view, which assumes that hadrons take
a certain shape under certain conditions of interaction. Following this
idea and taking into account the short-range nature of strong
interactions, the shape of hadrons and the region of their interaction
are mutually determined. In particular, the transverse sizes of hadrons
must be consistent with the transverse sizes of the interaction region,
and v.v. Similarly one should approach the question of the longitudinal
sizes.
Currently, it is widely believed that the longitudinal sizes of the
hadron interaction region grow linearly with the energy
\cite{Feinberg,Gribov65,Low,Schrempp,Gribov73,Petrov-Ryutin}. So, if
this is the case, then one should assume that the longitudinal sizes of
fast hadrons grow linearly with the energy, as well.

In fact, however, the latter assumption seems paradoxical, since
starting at certain energies the longitudinal sizes should reach the
macroscopic values, which is difficult to accept. As far as we know,
this issue has not been discussed anywhere. Apparently, by default it
was assumed that upon reaching certain energies further growth should
be stopped. However, no ideas were proposed for the implementation of
this scenario.

In this article, we develop an idea that the reason for limiting the
longitudinal growth of hadrons is the same that leads to the accelerated
expansion in the transverse directions at ultra-high energies. The
underlying reason is the emergence of the correlations between the
partons at ultra-high energies, which ultimately is related to the
confinement phenomenon. Moreover, for the same reason a hollow appears
inside the hadrons along the central axis in the direction of their
motion. We make estimates of the corresponding parameters in the case of
nucleons. In this regard, below we refer to nucleons, although the
qualitative results can be applied to other hadrons, as well.

The structure of the paper is as follows. In section \ref{sec2}, we give
necessary definitions and describe the evolution of the shape of
nucleons in the transverse projection. Section \ref{sec3} considers the
evolution in the longitudinal direction. The results are
discussed in section \ref{sec4}.

\section{Fast nucleons in the transverse projection}\label{sec2}

We proceed from the idea that nucleon at rest can be represented in
the first approximation as a bound state of three light quarks. Of 
course, the quarks can virtually split into other quarks and gluons, but
they immediately recombine back, so that the average number of quarks in
the nucleon at rest is approximately three. This can be understood as
follows. Each initial quark in the nucleon is distributed with a
variance that can be estimated as the radius $R_{\mbox{\scriptsize
v}}$ of the nucleon at rest. In turn, if quarks are distributed in the
configuration space, they are distributed in the momentum space, as
well, with a variance of $\mu \simeq 1/R_{\mbox{\scriptsize v}}$, which
is roughly 0.3 GeV. Furthermore, since the light quarks are almost
massless, their average energy and average momenta squared are connected
via the relation $\langle E^2 \rangle \simeq \langle p^2 \rangle$.
However, the average quark momentum is zero. So, $\langle p^2 \rangle$
is essentially the momentum variance. Consequently, $\langle E^2 \rangle
\simeq \mu^2$ and $\mu$ can be interpreted as the effective quark mass
in the nucleon. Owing to its large value, the splitting of quarks in the
nucleon at rest is strongly suppressed.

A completely different picture occurs in a fast-moving nucleon. Because
of the excess energy, the fast quarks and gluons---hereinafter referred
to as partons---split freely, acquiring effective (``transverse'') mass
due to their transverse motion. Simultaneously, they lose the
longitudinal momentum and the energy, until they become slow. When they
become slow, they can no longer split due to their large effective mass,
and the development of the decay cascade is completed. 

In the above processes, an important feature is that each splitting is
accompanied by a shift in the position and transverse momentum of newly
formed parton relative to the parton-parent. In the configuration space,
this leads to the shift of the partons during splitting to the
periphery. With increasing energy of the nucleon the decay chain becomes
longer, and the last partons in the cascades become more and more
distant from the geometric center. Correspondingly, the transverse size
of the nucleon increases with the energy. In the case of uncorrelated
partons motion the increase is proportional to $\sqrt{\,\ln s}$ (the
``diffusion motion'') \cite{Gribov73} and in the case of correlated
motion it may be proportional to $\ln s$ \cite{Gribov73009}. The latter
option was confirmed in \cite{Nekrasov}. Moreover, \cite{Nekrasov}
showed that in the former case the transverse root-mean-square (RMS)
momenta of the last partons in the cascades also grow as $\sqrt{\,\ln
s}$, but in the latter case they are limited.

Let us recall the basic definitions of \cite{Nekrasov}. We define the
effective transverse radius ${\cal R}(s)$ of the nucleon as the average
variance of the distance between the peripheral partons and the central
axis in the direction of motion of the nucleon. Peripheral partons are
those that arise at the end of the cascades, and they have the greatest
variance among the partons in cascades. It should be emphasized that the
number of patrons in the cascades is not fixed (the cascades may be
incomplete) and is determined by a certain probability distribution. The
above-mentioned average of the variance is determined by this
probability distribution. Similarly, we define the size ${\cal K}(s)$ of
the nucleon in the transverse momentum space as the average variance of
the transverse momenta of the peripheral partons. The latter partons
also have the largest RMS transverse momenta relative to the geometric
center of the nucleon.

\begin{flushleft}
{\bf The mode of uncorrelated motion of the partons}
\end{flushleft} 

One of fundamental results of \cite{Nekrasov} is the observation that in
the mode of uncorrelated (``diffusive'') motion of the partons both
${\cal R}$ and ${\cal K}$ are proportional to the average number
$\bar{N}$ of partons in cascades, 
\begin{equation}\label{s1}
{\cal R}^2 = \bar{N} \mu^{-2} \,, 
\end{equation} 
\begin{equation}\label{s2}
{\cal K}^2 = \bar{N} \mu^{2} \,.
\end{equation} 
Here $\mu$ is the variance of the transverse parton momentum relative to
the parent parton. This quantity is independent of the longitudinal
momentum of the nucleon, and coincides with that in the nucleon at rest
(the latter is obvious when $\bar{N}=1$). Let us emphasize once again,
that (\ref{s1}) and (\ref{s2}) include average values in the cascades.
However, the same relations are valid for the corresponding maximum
values, $N_m$, ${\cal R}_m^2$ and ${\cal K}_m^2$, related to the fully
developed cascades.

Two important consequences may be immediately deduced from (\ref{s1})
and (\ref{s2}). The first one is that the behavior of ${\cal R}^2$ and
${\cal K}^2$ is fully determined by the average number of the partons
$\bar{N}$. In the parton model $\bar{N}$ increases asymptotically as
$\ln s$ \cite{Gribov73009,Gribov73}. So ${\cal R}^2$ and ${\cal K}^2$
should increase with the energy as $\ln s$. The second consequence is
that the average parton density $\varrho$ in the transverse projection
is a constant, independent of the nucleon energy,
\begin{equation}\label{s3}
\varrho \equiv 3 \bar{N}/S = 3 \mu^2/ \pi \,.
\end{equation} 
Here $S=\pi {\cal R}^2$, and 3 is the number of the cascades in a
free-moving nucleon.\footnote{Actually, the number of cascades is
determined by the number of valence quarks, if the development of
cascades goes without branches. The latter property is confirmed at high
energies in the leading approximation, see e.g.~\cite{Gribov73009}. At
low energies, this is our assumption. If branching nevertheless occurs
at low energies, then 3 should be replaced by another larger number,
which should no longer change with increasing energy at high energies.}
The fact that $\varrho$ is constant means that nucleons can expand
unrestrictedly with increasing the energy as new partons are formed,
without any contradiction to the confinement.

Parameter $\mu$ must be determined more precisely. The problem is that
the sizes of the nucleons in strong and electromagnetic interactions,
differ from each other. Recently \cite{Petrov-Okorokov} defined the
transverse radius of strong-interacting valence quark core of a nucleon
as ${\cal R}_{\mbox{\scriptsize v}}^2 = \langle b_0^2 \rangle / 2$,
where $\langle b_0^2 \rangle$ is the impact parameter at
$s\!=\!s_0$, and $s_0$ is the energy at which the colliding nucleons
cease to overlap with each other in the impact parameter plane.
Phenomenologically, $s_0$ is the energy at which the growth rate of
$B(s)$ changes from very fast to logarithmic. So at $s \ge s_0$ (but not
too large $s$) the slope parameter is 
\begin{equation}\label{s4}
B(s) = B_0 + 2 \alpha'_0 \ln (s/s_0)\,,
\end{equation} 
where $\alpha'_0$ is a phenomenological parameter. The estimates of
\cite{Petrov-Okorokov} are as follows: $\sqrt{s_0} \approx
10$ GeV, $B_0 \equiv B(s_0) = 11.10 \pm 0.26$ GeV$^{-2}$, and
in view of $B_0 = \langle b_0^2 \rangle / 2$, ${\cal
R}_{\mbox{\scriptsize v}} = \sqrt{B_0} = 0.656 \pm 0.008$ fm. Notice
that in this case the three-dimensional radius of the valence quark core
is $R_{\mbox{\scriptsize v}}^2 = (3/2) {\cal R}_{\mbox{\scriptsize v}}^2
= (0.805 \pm 0.009 \, \mbox{fm})^2$.

We generalize the above approach by supplementing it with the
assumption that for any $s \ge s_0$ the actual transverse size of a
nucleon is determined by the relation ${\cal R}^2 = \langle b^2 \rangle
/2$. So, given $\langle b^2 \rangle / 2 = B$, we have 
\begin{equation}\label{s5}
{\cal R}^2(s) = {\cal R}_0^2 + 2 \alpha'_0 \ln (s/s_0) \,,
\end{equation} 
where ${\cal R}_0^2 = B_0$. In fact, (\ref{s5}) is a realization of the
concept that the region of interaction between nucleons must be
comparable with their actual sizes. Reversing the logic, we can say that
with increasing the energy the radii of nucleons increase and,
therefore, the RMS impact parameter increases.

It should be emphasized that in our approach ${\cal
R}_{\mbox{\scriptsize v}}$ differs from ${\cal R}_0$. To get ${\cal
R}_{\mbox{\scriptsize v}}$, we must go down in energy to the nucleon at
rest, i.e.~to $\sqrt{s}/2 = M$ where $M$ is the nucleon mass. The basis
for this extrapolation is that relation (\ref{s5}) is a consequence of
the diffusion process which starts from the rest state of the nucleon.
(In contrast, formula (\ref{s4}) is inapplicable at $s < s_0$, since at
these energies nucleons are scattering in the overlap mode of their
physical sizes.) Taking this into account, we get
\begin{equation}\label{s6}
{\cal R}_{\mbox{\scriptsize v}}^2 = {\cal R}_0^2 + 2 \alpha'_0 \ln
(4 M^2/s_0) \,.
\end{equation} 
With $\alpha'_0 = 0.25$ GeV$^{-2}$ \cite{Amaldi}, this leads to ${\cal
R}_{\mbox{\scriptsize v}} = 0.60$ fm ($R_{\mbox{\scriptsize v}} = 0.74
\, \mbox{fm}$). Hence we have $\mu = 1/{\cal R}_{\mbox{\scriptsize v}} =
0.33$ GeV and $\varrho = (0.62 \, \mbox{fm})^{-2}$. Note that ${\cal
R}_{\mbox{\scriptsize v}}$ is weakly dependent on $\alpha'_0$. For
example, $\alpha'_0 = 0.321$ GeV$^{-2}$ \cite{DL} gives ${\cal
R}_{\mbox{\scriptsize v}} = 0.59$ fm and, within the accepted accuracy,
the same value for $\mu$. Below, we do estimates with $\alpha'_0 = 0.25$
GeV$^{-2}$ and $\mu = 0.33$ GeV.

Now we turn to other formulas. In view of (\ref{s1}) and (\ref{s5}) we
have
\begin{equation}\label{s7}
\bar{N} = \mu^2 \left[{\cal R}_0^2 + 2 \alpha'_0 \ln (s/s_0) \right] .
\end{equation}
This gives $\bar{N}_{\mbox{\scriptsize v}}=1$ at $\sqrt{s}=2M$, as
required. An important characteristic is the maximum number $N_m$ of
partons in cascades. At high energies in the multi-peripheral picture
$N_m = \gamma \ln (2 P/\mu)$, where $\gamma \simeq 1/\!\ln 2$ and $P$ is
the longitudinal momentum of the cascade \cite{Gribov73009,Gribov73}. In
the case of nucleon at rest, $P$ makes sense of the RMS momentum of the
valence quarks ($P = \mu$ leads to $N_m = 1$). In the case of a fast
nucleon, assuming its momentum is equally distributed between the
cascades, we have $P = \sqrt{s}/6$. This gives
\begin{equation}\label{s8}
N_m \simeq \gamma \ln \frac{\sqrt{s}}{3\mu} \;.
\end{equation} 
(This approximation also works well for the nucleon at rest, since
$\sqrt{s} = 2M$ gives $N_m \simeq 0.94$, which is close to 1.) Based on
(\ref{s7}) and (\ref{s8}), we define the parameter of saturation of the
cascades, $\varkappa = \bar{N} / N_m$. It monotonically decreases from 1
at $\sqrt{s} = 2M$ to $4\alpha'_0 \, \mu^2 / \gamma \simeq 0.07$ at $s
\to \infty$. 

Finally, in view of (\ref{s2}), (\ref{s6}) and (\ref{s7}), the RMS
transverse momentum of the peripheral parton is
\begin{equation}\label{s9}
{\cal K}^2(s) = \mu^2 \left[ 1 + 2 \alpha'_0 \,\mu^2 \ln (s/4M^2)
\right] .
\end{equation}
Obviously, the growth of ${\cal K}^2$ with $s$ should not continue
indefinitely, since too large transverse motions would lead to problems
with integrity of the nucleon. So, the ``diffusive'' mode should end at
some energies. It is worth noting that the maximum momentum in the
cascades ${\cal K}_m$ is more critical in this sense, since the growth
of ${\cal K}_m$ exceeds that of ${\cal K}$ in view of ${\cal K}_m =
{\cal K} / \sqrt{\varkappa}$.

Table \ref{T1} shows the values of the above quantities for various $s$,
starting from $s=s_0$ and further including the energies of ISR,
Sp$\bar{\mbox p}$S, Tevatron, and LHC. For illustration, we also show
results for $\sqrt{s}=100$ TeV. We note that ${\cal K}$ gives the lower
estimate for the average transverse momenta of secondary particles in
the inelastic processes. The mentioned averages are as follows: 0.4 GeV
at $\sqrt{s} = 31.5$--63 GeV \cite{ISR}, 0.42 GeV at $\sqrt{s} = 540$
GeV \cite{SppS}, 0.50 GeV at 2.36 TeV \cite{Khachatryan1}, 0.55 GeV at 7
TeV \cite{Khachatryan2}. So, the limited growth of ${\cal K}$ with the
energy is quite consistent with the data. 

\begin{table*}
\caption{Estimated parameters in the mode of uncorrelated motion of the
partons: the effective transverse radius of the nucleon ${\cal R}$;
transverse RMS momenta of the peripheral partons, average ${\cal K}$ and
maximum ${\cal K}_m$ in cascades; the number of the partons in cascades,
average $\bar{N}$ and maximum $N_m$; cascade saturation parameter
$\!\varkappa$.
\vspace*{0.3\baselineskip} 
}\label{T1}
\begin{tabular*}{\textwidth}{@{\extracolsep{\fill}}lllllll@{}}
\hline\noalign{\medskip}
 $\sqrt{s}$ {\small [GeV]} & ${\cal R}$ {\small [fm]} 
& ${\cal K}$ {\small [GeV]} & ${\cal K}_m$ {\small [GeV]} & $\bar{N}$ &
$N_m$ & $\varkappa$ 
\\[1mm]
\hline\noalign{\medskip}
10     & 0.66 & 0.35 & 0.60 & 1.18 & 3.36 & 0.35
\\[1mm]
50     & 0.70 & 0.38 & 0.78 & 1.35 & 5.68 & 0.24
\\[1mm]
500    & 0.76 & 0.41 & 0.98 & 1.59 & 9.00 & 0.18
\\[1mm]
2000   & 0.80 & 0.43 & 1.08 & 1.74 & 11.00 & 0.16
\\[1mm]
7000   & 0.83 & 0.45 & 1.17 & 1.87 & 12.81 & 0.15
\\[1mm]
13000  & 0.84 & 0.45 & 1.21 & 1.94 & 13.70 & 0.14
\\[1mm]
100000 & 0.89 & 0.48 & 1.33 & 2.15 & 16.64 & 0.13
\\
\noalign{\smallskip}\hline
\end{tabular*} 
\end{table*} 

\begin{flushleft}
{\bf The mode of correlated motion of the partons}
\end{flushleft}

The fundamental feature in the mode of correlated motion is a fixed
value of ${\cal K}$ and, at the same time, an accelerated growth of
${\cal R}$, and in such a way that the uncertainty relation for cascade
processes is satisfied \cite{Nekrasov}. In the notation of this work,
the latter relation is
\begin{equation}\label{s10}
{\cal R} \, {\cal K} \ge \bar{N}  \,.
\end{equation} 
So, if for $s=s_1$ the grows of ${\cal K}$ stops at a certain ${\cal K}
= {\cal K}_1$,\footnote{Recall that limiting the RMS momentum does not
prohibit its large fluctuations, but means their low probability.} then
${\cal R}$ continues evolution for $s > s_1$ in accordance with the
relation
\begin{equation}\label{s11}
{\cal R}  \ge \bar{N}/{\cal K}_1   \,.
\end{equation} 

Let us now take into account that $B(s) \sim \ln^2 \!s$ at ultra-high
$s$. Assuming that the change in the behavior of $B(s)$ occurs at
$s=s_1$, we get for $s \ge s_1$
\begin{equation}\label{s12}
\widehat{B} (s) = B_1 + A \ln^2 (s/s_1)\,.
\end{equation} 
Here $A$ is a parameter, $B_1 = B(s_1)$ provides crosslinking with
(\ref{s4}) at $s=s_1$. Symbol ``{\Large $\;_{\widehat{}}\;$}'' at the
top means that the corresponding quantity relates to the mode $s \ge
s_1$.

Next, we apply the provision that the size of strongly interacting
nucleons is half the impact parameter in quadrature. Given
$\langle \,\widehat{b}\,^2 \rangle / 2 = \widehat{B}$, this leads to
\begin{equation}\label{s13}
\widehat{\cal R}^2(s) = {\cal R}^2_1 + A \ln^2 (s/s_1)\,,
\end{equation}
where ${\cal R}^2_1 = B_1$. In view of (\ref{s11}) and (\ref{s13}) we
also have
\begin{equation}\label{s14}
\widehat{\bar{N}} \simeq {\cal K}_1 \left[
{\cal R}^2_1 + A \ln^2 (s/s_1) \right]^{1/2} .
\end{equation}
This means $\widehat{\bar{N}} \sim \ln s$ as $s \to \infty$. The
crosslinking between (\ref{s14}) and (\ref{s7}) at $s = s_1$ is
automatic if (\ref{s10}) implies equality. However, the equality is not
necessary if the mode change occurs in a finite region. 

The consequence of (\ref{s13}) and (\ref{s14}) is a fundamentally
different behavior of the average parton density in the transverse
projection,
\begin{equation}\label{s15}
\widehat{\varrho} = (3 {\cal K}_1 / \pi) \left[
{\cal R}^2_1 + A \ln^2 (s/s_1) \right]^{-1/2}\,.
\end{equation}
Recall that in the uncorrelated (``diffusion'') mode $\varrho$ is an
energy independent constant. Unlike that, $\widehat{\varrho}$ decreases
with increasing the energy. This implies that each parton occupies more
and more space in the impact parameter plane and, therefore, the
distance between the partons increases indefinitely. Obviously, this is
incompatible with the confinement. The only reasonable way to resolve
the contradiction is to assume the appearance of an expanding hollow
inside the nucleons, mostly free of partons. Due to the symmetry, the
hollow is located along the geometric central axis of the nucleon. So,
the nucleon in a rough approximation takes the form of a ring in the
transverse projection.

The hollow starts forming when the density reaches a critical (small)
value. Formula (\ref{s15}) does not work in the presence of the hollow,
but we can express its radius $r$ through the local density
$\widehat{\varrho}$. Based on the definition
$\widehat{\varrho} = 3 \widehat{\bar{N}} / S$, $S = [\pi (\widehat{{\cal
R}} {^2} \!-\! r^2) ]$, we get
\begin{equation}\label{s16}
\widehat{r}^2 = \widehat{\cal R} \left( \widehat{\cal R} - \nu
{\cal R}_1 \right) ,
\end{equation} 
where $\nu = \varrho / \widehat{\varrho} \ge 1$. In the limiting case
$\nu = 1$, the hollow starts to form at $s = s_1$. In the general case,
the hollow appears when the outer radius $\widehat{\cal R}$ of the
nucleon reaches $\nu {\cal R}_1$. The thickness of the ring at this
moment is maximum (equals the outer radius), and then it monotonically
decreases to $\nu {\cal R}_1 / 2$ with increasing $s$. In any case $r$
grows asymptotically as $\ln s$, following $\widehat{\cal R}$ minus the
thickness of the ring. 

Table \ref{T2} shows numerical estimates in the case $\nu=1$ at
$\sqrt{s_1} = 2$ TeV and $\sqrt{s_1} = 7$ TeV (the results are separated
by ``/''). Based on the fit \cite{Schegelsky-Ryskin}, we use $A =
0.029$. 

In conclusion, we note that it is a priori unclear whether the hollow
inside the nucleons affect the region of their interaction. However, the
idea of the appearance at high energies of a hollow in the nucleon
interaction region is intensively discussed, see
\cite{Troshin2007,Troshin2019,Dremin2015} and the references therein.
Irrespective of this, recent analysis \cite{Dremin2020} of the TOTEM
data at 13 TeV led to the conclusion about the multilayer structure of
the protons, with the spatial size of the inner layer near 0.45 fm. This
value should be compared with $r$ in Table \ref{T2}.

\begin{table*}
\caption{Estimated parameters in the mode of correlated motion at
$\sqrt{s_1} = 2$ TeV / 7 TeV. The notation is similar to
that of Table \ref{T1}. }\label{T2}
\vspace*{0.3\baselineskip} 
\begin{tabular*}{\textwidth}{@{\extracolsep{\fill}}llllll@{}}
\hline\noalign{\smallskip}
 $\sqrt{s}$ {\small [GeV]} & $\widehat{\cal R}$ {\small [fm]} 
& $r$ {\small [fm]} & $\widehat{\bar{N}}$ & $N_m$ 
& $\widehat{\varkappa}$
\\[1mm]
\hline\noalign{\medskip}
2000  & 0.80        & 0.          & 1.74        & 11.00 & 0.16
\\[1mm]
7000  & 0.90 / 0.83 & 0.31 / 0.   & 1.97 / 1.87 & 12.81 & 0.15 / 0.15
\\[1mm]
13000 & 1.02 / 0.85 & 0.48 / 0.15 & 2.23 / 1.93 & 13.70 & 0.16 / 0.14
\\[1mm]
100000& 1.55 / 1.23 & 1.08 / 0.70 & 3.39 / 2.78 & 16.64 & 0.20 / 0.17
\\
\noalign{\smallskip}\hline 
\end{tabular*} 
\end{table*}

\section{The longitudinal size}\label{sec3}

Above, we have identified two features that indicate a change in the
mode of motion of the partons. One of them is phenomenological (the
change in the behavior of the slope of the diffraction cone), and the
other is the internal inconsistency (unrestricted growth of the RMS
transverse momenta in the mode of uncorrelated motion). However, the
fundamental reason for the mode change remains unknown. We will see
below that it is related to the longitudinal size of nucleons.

We mentioned in the Introduction the widespread opinion about the
longitudinal region of interaction of fast nucleons, that it increases
linearly with increasing energy of the collision
\cite{Feinberg,Gribov65,Low,Schrempp,Gribov73,Petrov-Ryutin}. From the
viewpoint of the parton model this can be explained by the fact that
nucleons interact in collisions mainly through slow partons, and the
time of development of quantum fluctuations leading to slow partons is
proportional to the energy of the nucleons \cite{Gribov73009,Gribov73}.
For this reason the variance of the hadron interaction region in the
longitudinal direction increases linearly with the energy. Now we recall
that the interaction region of slow partons is of the order of $1/ \mu$.
Hence it follows that the longitudinal sizes of nucleons and of their
interaction region must coincide with the $1/\mu$-accuracy. Based on
this, we conclude that the longitudinal size of the nucleons increases
linearly with their energy.

The next question is how far the longitudinal size can grow. To answer,
we relate the number of partons in the rapidly moving nucleon to its
three-dimensional volume. We have seen above that the number of partons
increases logarithmically with the energy, the cross-section area also
increases logarithmically (in the mode of uncorrelated motion of
partons) and the longitudinal size of the nucleon increases, too. This
means that the three-dimensional density of the partons decreases with
increasing energy, i.e.~partons become less densely packed in the
nucleon. Obviously, this should lead to an increase in the strong
coupling between the partons. We regard this fact as a strict indication
for establishing the mode of their correlated behavior.

In this mode, the nucleons interact in collisions not through slow
(uncorrelated) partons, but as integral objects. Moreover, the slow
partons, which are non-relativistic, are not formed at all in this
mode.\footnote{This should not lead to significant distortion of
asymptotic estimate (\ref{s8}), because only a limited number of partons
are cut off.} The reason is actually indicated above. Namely, the
appearance of slow partons leads to an increase in the length of the
nucleon and, hence, to a decrease in the density of the partons. The
latter property leads to an increase in the coupling between the
partons, and this prevents the appearance of non-relativistic slow
partons which inevitably lag behind the bulk of other (relativistic)
partons during the longitudinal motion. Practically, this means stopping
the growth of the longitudinal size of nucleons.

The simplest model that parametrically describes the evolution of the
longitudinal size of nucleons looks as follows. We use a purely
geometric interpretation, i.e.~we assume that the nucleon at rest can be
represented as a regular ball with a three-dimensional radius
$R_{\mbox{\scriptsize v}}^2 = (3/2) {\cal R}_{\mbox{\scriptsize v}}^2$.
In this case the linear growth of the longitudinal radius $R_L $ of the
nucleon is written as
\begin{equation}\label{s17}
R_L = R_{\mbox{\scriptsize v}} 
\left[ 1 + \xi \left(\frac{E}{M} -1 \right) \right] .
\end{equation} 
Here $E = \sqrt{s}/2$ is the energy of the nucleon, $M$ is its mass, and
$\xi$ is a parameter. We estimate $\xi$ from the condition that 
three-dimensional density of the partons reaches a critical (small)
value $\rho_c$. Namely, assuming the spatial volume of the fast nucleon
is $2 R_L \pi {\cal R}^2$, we have $\rho(s) = \varrho / (2 R_L)$, where
the two-dimensional density $\varrho$ is given in (\ref{s3}). The
critical density $\rho_c$ can be estimated as the inverse volume $V_c =
(4/3)\pi r_c^3$, where $r_c$ (``confinement radius'') determines the
maximum distance by which the partons can move away from each other. The
condition $\rho(s_1) = \rho_c $ gives 
\begin{equation}\label{s18}
2 {R_L}_{\lvert_{E=E_1}} = V_c \, \varrho \,,
\end{equation} 
whence we get $\xi$. The result is very sensitive to $r_c$. In
particular, assuming $r_c=1$ fm (should be compared with
$R_{\mbox{\scriptsize v}} = 0.74$ fm), we get $\xi = 0.0060$ at
$\sqrt{s_1} = 2$ TeV and $\xi = 0.0017$ at $\sqrt{s_1} = 7$ TeV.
Assuming $r_c=1.2$ fm, we get $\xi = 0.0110$ and $\xi = 0.0032$,
respectively.

Table \ref{T3} shows the results for $R_L$ in the case $r_c=1$ fm at
$\sqrt{s_1} = 2$ TeV and $\sqrt{s_1} = 7$ TeV. Recall that for $s > s_1$
the growth of $R_L$ stops, and the three-dimensional local density of
the partons is constant. 

\begin{table*}
\caption{The longitudinal radius $R_L$ of the nucleon.}\label{T3}
\vspace*{0.5\baselineskip} 
\begin{tabular*}{\textwidth}{@{\extracolsep{\fill}}llllll@{}}
\hline\noalign{\medskip}
\ $\sqrt{s}$ {\small [GeV]} & 10   & 50    & 500    & 2000 & 7000 \
\\
\hline\noalign{\medskip}
\\[-4mm]
\ \ $R_L$ {\small [fm]} \ {\footnotesize ($\sqrt{s_1} = 2$ TeV)}
& 0.76 & 0.85  & 1.92   & 5.47 & 5.47 
\\[2mm]
\ \ $R_L$ {\small [fm]} \ {\footnotesize ($\sqrt{s_1} = 7$ TeV)} 
& 0.75 & 0.77  & 1.08   & 2.77 & 5.47 
\\
\noalign{\smallskip}\hline
\end{tabular*} 
\end{table*}

\section{Discussion}\label{sec4}

We have investigated the shape of free moving fast nucleons (in the
limit $t=0$ in elastic collisions, excluding electromagnetic
interactions).
The analysis has been carried out in the extended parton model, which
consistently takes into account the transverse motions of the partons.
The choice of the parton model as a research tool is caused by
impossibility to solve the problem by means of perturbative QCD. At the
same time, the extended parton model makes it possible to understand the
behavior of partons from the point of view of field theory without a
detailed description of the dynamics of their interactions. The idea of
expanding the parton model by means of taking into account the
transverse motions was formulated in \cite{Gribov73} and then developed
in \cite{Nekrasov}. In the given work to solve the aforementioned
problem, we use additional ideas external to the parton model. 

The general idea is based on the assumption that the sizes of nucleons
are comparable to the sizes of their interaction region. Physically,
this is caused by the short-range property of strong interactions. In
the case of transverse directions, this is strongly supported by the
observation that the mean-square transverse radius ${\cal R}^2$ of the
nucleons, predicted in the parton model, grows with the energy in the
same way as the transverse mean-square distance $\langle b^2 \rangle$
between the centers of their collision. This allows us to postulate
equality ${\cal R}^2 = \langle b^2 \rangle /2$ which is to be satisfied
for any $s \ge s_0$, where $s_0$ is the energy above which the colliding
nucleons cease to overlap with each other in the impact parameter plane
($\sqrt{s_0}$ is estimated as 10 GeV \cite{Petrov-Okorokov}). Similarly,
the longitudinal sizes of the nucleons should correlate with the
longitudinal sizes of their interaction region. In particular, if the
longitudinal size of the interaction region grows linearly with the
energy, then the longitudinal sizes of the nucleons must also grow
linearly, and vice versa.

Another additional point is the assumption that the coupling between the
partons increases as their spatial density decreases. We interpret this
increase as a reason for the transition to the mode of correlated motion
of the partons at the energy above a certain $s_1$ ($s_1 > s_0$). In
this mode the transverse sizes of the nucleon grow faster than in the
uncorrelated mode, but the RMS transverse momenta of the partons become
limited. Concurrently, a hollow is formed inside the nucleon with a
radius growing in proportion to its outer transverse radius. The hollow
prevents a decrease in the local density of partons with the accelerated
growth of the outer radius. The longitudinal size of the nucleon cannot
grow in this mode, since the correlated behavior of the partons does not
imply the presence of slow (non-relativistic) partons.

The quantitative estimates made in this article are mainly illustrative.
However, they give an idea of the scale of the parameters and the trend
of their change with the energy. For more accurate estimates a more
detailed dynamic model is needed. Unfortunately, it was not possible to
evaluate the basic parameter $s_1$ of the transition from the one mode
to another, but this is a common difficulty in determining the
transition point from  $\ln s$ to $\ln^2 \!s$. 

An important conclusion that can be drawn from the above results is that
the longitudinal sizes of the nucleons have a decisive influence on
their observable properties (despite the fact that the longitudinal
sizes themselves are not directly observable). The reason lies in the
fundamental dependence of the spatial density of partons on the
longitudinal size of the nucleon. When the density drops to a critical
value, the mode of motion of the partons changes from uncorrelated to
correlated. The observed manifestations of this transition are
the changes in the behavior of the slope of the diffraction cone and of
the total cross section. The increase in the coupling between the
partons accompanying the transition can also serve as a qualitative
explanation for the observed \cite{Antchev} increase with the energy in
the ratio of the elastic to inelastic cross section. The latter issue
will be discussed elsewhere.

On the whole, the above results provide a better understanding of the
processes occurring in the soft elastic collisions of nucleons at high
energies, and can serve as a guide for further studies.

\bigskip

The author is grateful to R.A.Ryutin for useful discussions.

\end{document}